Working Paper October 3, 2026

# New and Improved Analytical Methods for Calculating Power System Adequacy Metrics

Chanan Singh
University Distinguished Professor
Department of Electrical and Computer Engineering
Texas A&M University

## Abstract

A new analytical technique is proposed for calculating  adequacy metrics for generation planning. The proposed technique is simple, direct, and provides higher accuracy than traditional discretized approaches. Additionally, a new analytical methodology is introduced for calculating loss of load days (LOL Days) as distinct from loss of load events or frequency of load loss. This paper emphasizes that, where feasible, analytical methods are preferable to Monte Carlo simulations because they yield exact closed-form solutions rather than statistical estimates.

## Introduction

Probabilistic resource adequacy metrics have been proposed and successfully applied to power systems for several decades [1,2,3]. With the advent of digital computers, highly sophisticated techniques were developed for their computation. Historically, these metrics were calculated using the load and generation models to determine the adequate level of installed reserve, assuming a generation fleet composed primarily of fossil-fuel-fired thermal or hydroelectric units.

More recently, there has been a renewed interest in probabilistic resource adequacy due to key industry drivers, including the rapid expansion of variable energy resources (VERs) like wind and solar, and the massive load growth associated with data centers. Although several comprehensive textbooks and papers exist on this topic [1, 2, 3], many practitioners and researchers currently involved in probabilistic evaluation lack formal education in classical power system reliability. This has led to the so-called "reinvention of the wheel" and unnecessary renaming of established metrics. For example, the well-established index *Loss of Load Frequency* (LOLF) has recently been renamed in some reports as *LOLEV* (Loss of Load Events) [4]. While there is nothing fundamentally incorrect with this renaming, it creates avoidable confusion in the literature nationally and internationally.

Nonetheless, the definitions of certain indices have undergone legitimate modifications to reflect changes in the modern generation mix [4]. For instance, the flattening of daily load curves due to demand response and the high-frequency fluctuations of solar and wind generation mean that traditional peak-load approximations are no longer sufficient.

The primary objective of this paper is to present improved and mathematically rigorous analytical methods for calculating these critical indices.

## Reliability/Adequacy Metrics and modifications

The various indices proposed for resource adequacy are briefly explained below.

1. **Loss of Load Expectation (LOLE)**
   Loss of load expectation can be based on daily peaks or hourly loads.

   a. DLOLE or LOLE is the expected number of days per year on which insufficient generating capacity is available to serve the daily peak load. Originally it was based on the loss during peak load as the peak load was significantly higher than load during the rest of the hours. So, the logic was that if the peak load were served then the rest of the hours would be served as well. Later some utilities replaced the daily peak by load loss during any hour of the day. The reason for this is that load curve has tended to flatten because of demand response and sources like wind and solar tend to fluctuate more frequently.

   b. HLOLE or LOLH is the expected number of hours per year when insufficient generating capacity is available to serve the load. Please note that LOLH is the total number of hours of load loss and this index does not give any information about the mean of the event duration.

   c. The LOLE index gives no information on several important system reliability attributes:
      - Magnitude of capacity shortages when they occur.
      - Duration of capacity shortage events. HLOLE gives the mean hours of loss of load in a year but not the mean duration of load loss events that comprise LOLH.
      - Expected amount of unserved energy

2. **Frequency and Duration of Capacity Shortage Events (F&D)**
   In the frequency and duration approach following indexes are calculated.

   a. Frequency of generating capacity shortage events is defined to be the expected (mean) number of such events per year. This metric of frequency is also called LOLEV [4].
   b. Duration is the expected length of capacity shortage periods when they occur.
   c. F&D indices use hourly load information and thus reflect the influences of daily load cycle shape.
   d. F&D methods model unit parameters more fully than those models used in LOLE. These indices need the transition rate between various states of generators whereas the LOLE and EUE indices need only the probabilities of various states of generators.
   e. F&D indices are conceptually superior to LOLE. They have, however, greater data requirements.
   f. It should be noted that LOLH is the product of the frequency of load loss and its mean duration.

3. Expected Unserved Energy (EUE)
    a. The EUE index measures the expected amount of energy which will fail to be supplied per year due to generative capacity differences and/or shortages in basic energy supplies.

## Computation of Adequacy Metrics

The methods for adequacy or reliability computations can be broadly classified into analytical approaches and Monte Carlo simulation. The analytical approaches give one unique answer for adequacy metrics within the assumptions made whereas the Monte Carlo provides a statistical estimate of the value of these metrics. If we make resources adequacy calculations without considering transmission limitations, then analytical methods are very efficient and give exact answers. For looped transmission consideration, Monte Carlo provides more flexibility. Here we focus on the analytical methods.

There are several methods described for resource adequacy in the literature but in our opinion the ones described in references 3 and 5 are the most comprehensive. These methods proceed in the following steps.

### Build Generation System Model

The process starts with multistate models of generators where each state is assigned a capacity and interstate transition rates are specified. The models of generators are combined using very efficient unit addition algorithm [3,5] to create the generation system model. The generation system model is described by the following vectors:

Ci = ith element of vector **C**

= one of the discrete capacity outage levels

Pi = ith element of **P**

= probability of capacity outage greater than or equal to Ci

Fi = ith element of **F**

= frequency of capacity outage greater than or equal to Ci

### Build Load Model

The load model is built by scanning the hourly loads in a sequential manner as shown in Figure 1. The load levels are discretized and for a given load level Li, the following parameters are calculated:

$P_l(Li), F_l(Li)$ = Probability and frequency of load greater than or equal to Li MW.

Li+1 - Li = Z, constant increment for discretization.

These quantities are calculated using (1) and (2)

$$P_l(Li) = \frac{Number\ of\ hours \geq Li}{Total\ number\ of\ hours\ in\ the\ interval} \quad (1)$$

$$F_l(Li) = \frac{Transitions\ from\ less\ than\ Li\ to\ \geq Li}{Hours\ in\ the\ interval} \quad (2)$$

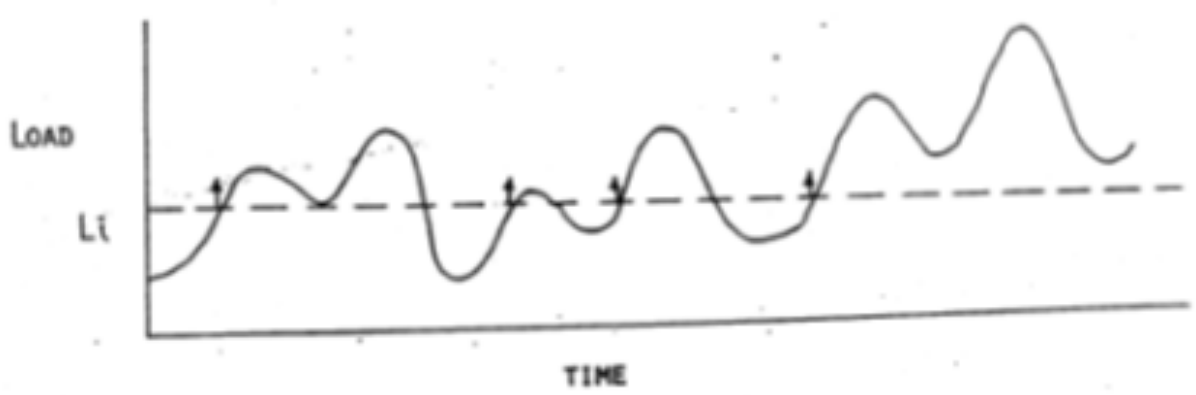

Fig 1. Hourly loads

It should be noted here that the sequential nature of the hourly load is fully modeled in this analytical approach. Notwithstanding this, Ref 4 incorrectly states, “It should be noted that LOLEV (and other event-based metrics) cannot be calculated explicitly using analytic convolution methods because those methods do not include temporal-dependent (i.e., chronological) information.” Perhaps the authors of ref 4 are not aware of the full capabilities of analytical methods described in [3].

**Build Generation Reserve Model**

The numerical difference between available generation and load is called the margin. The negative margins represent various depths of load loss. The cumulative probability of first negative margin is the probability of loss of load. So, the generation system model and load model are combined to determine the probability and frequency of a specified margin M. This is achieved using the equations 3-6.

$$\mathrm{P(M)} = \sum_{i=1}^{m} [P_i - P_{i+1}]\, P_l(C - C_i - M). \quad (3)$$

$$\mathrm{F(M)} = \mathrm{F}^g(\mathrm{M}) + \mathrm{F}^l(\mathrm{M}). \quad (4)$$

Where

$$F^g(M) = \sum_{i=1}^{m} [F_i - F_{i+1}]\, P_l(C - C_i - M) \quad (5)$$

$$F^l(M) = \sum_{i=1}^{m} [P_i - P_{i+1}]\, F_l(C - C_i - M) \quad (6)$$

P(M), F(M) = probability and frequency of margin less than or equal to M

$F^g(M)$, $F^l(M)$= components of F(M) due to generation and load changes respectively

C = installed capacity minus capacity on planned outage.

$C_i$ = capacity outage in state i

m= number of states in the generation system model.

### Calculation of Adequacy Metrics

The metrics of adequacy can now be calculated:

1. HLOLE

$$\text{HLOLE} = \text{P}(-\delta) \cdot \text{D}. \qquad (7)$$

where

$-\delta$ = negative margin with smallest absolute value

D = duration of the period of study

P($-\delta$) is the probability of $-\delta$ or less margin, i.e. the probability of all negative margins.

2. LOLE

The LOLE based on daily peak load can be computed by creating an hourly load curve with load for each hour equal to the daily peak load. Then the load model derived from this hourly load curve is convolved with the generation system model and

$$\text{LOLE} = \text{P}(-\delta)\ \text{D}/24 \qquad (8)$$

. 3. EUE

The expected unserved energy is calculated as,

$$\text{EUE} = \Delta\text{MD}\{\sum_{M=0}^{-L} P(M) - \frac{1}{2}[\text{P}(0) + \text{P}(-\text{L})]\}. \qquad (9)$$

where
ΔM = increment at which P(M) is computed
D = length of time interval in hours
-L = negative margin with largest absolute value (margins with probabilities less than $10^{-7}$ can be generally ignored)

### Proposed New Methods

Two new/improved methods are proposed. The first improved method is for calculation of daily based LOLE. The second new technique is for calculating the metrics without having to build the load model and then the generation reserve model. It convolves the generation system model directly with hourly load and cab thus avoid the need for discretization.

### Computing Loss of Load Event Days

Loss of load events or frequency is the expected number of events of loss of load. When the system comes out of the loss of load and then enters loss of load again, it encounters the second event. So, it is possible that there could be multiple loss of load events on a single day if the outage is long and the load is fluctuating. If we are calculating frequency or LOLEV, then these event should be treated separately. However, if we want to find Loss of Load Days, then these are the days on which a loss of load occurred once or multiple times. If we are interested to find the Expected Loss of Load Days, then all we need to do is modify the load model while using equation (2). In traversing hourly load in Fig 1, we need to identify the boundary of days. Then, we will count the number of days on which one or more transitions happened to make load greater or equal than Li. In other words, if there are multiple events within the boundary of the day, we count these as a single event.

### Direct interaction between Generation Model and Load Method

A new method is proposed to compute adequacy indices by combining generation system model with the hourly load directly without creating a load model or generation reserve model as is done in the current methods directly or indirectly. This method is also more accurate as it does not need any discretization of the load.

### LOLP and LOLH

Traversing each hour sequentially the loss of load probability for an hour can be computed by the product of probability of generation contributing to loss and the probability of an hour. Let us assume that the load for an hour i is Li. Then loss of load will occur if

( C-Ci) $\leq Li$

LOLP = $\sum_{1}^{N} Pi \frac{1}{N}$ (10)

Where

N is the number of hours in the interval, 8760 if the interval is a year.
Pi. Is the probability of capacity outage equal to or greater than Ci
Ci. is smallest capacity outage causing loss of load at hour i with load Li

On a yearly basis,
LOLH = LOLPx8760

### LOLF or LOLEV

Using the concept of frequency balance [3], steady state frequency of an event can be calculated either as the expected transitions out of the boundary defining the event or into the boundary. We will compute expected transitions out of the boundary defining the loss of load. Then for hour i,

Frequency of load loss for hour i due to generation= $Fi \frac{1}{N}$ (11)

Where
Fi is the frequency of outage equal to or greater than Ci
Ci is smallest capacity outage causing loss of load at hour i with load Li

1/N.    Probability of load for one hour

Frequency of load loss for hour i due to load change = (Pi – Pi+1) $\frac{1}{N}\,\alpha$.    (12)
Where $\alpha$ is 1 if (Pi – Pi+1) is positive and 0 otherwise.

The following discussion helps in understanding the equation (12). If the load goes up from hour i to i+1, then (Pi – Pi+1) is negative and so alpha will be zero. If the load decreases from i to i+1, then (Pi-Pi+1) is positive and represents the probability of generation states that transit from loss of load to no loss of load. If (Pi-Pi+1) is multiplied by transition rate of load from i to i+1, which is (1/one hour), that will yield the frequency due to load change. So

LOLF = $\sum_1^N Fi\,\frac{1}{N} + \sum_1^N (\mathrm{Pi} - \mathrm{Pi}+1)\mathrm{x}\frac{1}{N}\mathrm{x}\alpha$    (13)

## Conclusions

This paper has presented advanced analytical formulations for calculating power system resource adequacy metrics. The key contributions include:

1. Correcting misconceptions in recent literature regarding the ability of analytical convolution methods to model chronological load characteristics.
2. Introducing a modified load-scanning technique to analytically calculate Expected Loss of Load Days.
3. Proposing a direct, non-discretized method that combines the generation system model directly with chronological hourly loads, improving calculation speed and eliminating discretization errors.

These methods provide researchers and utilities with mathematically exact, closed-form solutions, offering a powerful alternative to computationally intensive Monte Carlo simulat

References;

1. John Endrenyi, Reliability Modeling in Electric Power Systems, Jon Wiley and Sons, 1978.
2. Roy Billinton and Ron Allan, Reliability Evaluation of Power Systems, Plenum Press, 1984
3. C. Singh, P. Jirutitijaroen, J. Mitra, Electric Power Grid Reliability Evaluation: Models and Methods, John Wiley, 2019.
4. NERC Draft Report, Probabilistic Adequacy and Measures, May 13, 2026
5. C.Singh, Lecture notes on Power System Reliability, https://chanansingh.engr.tamu.edu